\documentclass[conference,final,twoside]{IEEEtran}
\usepackage{cite}

\ifCLASSINFOpdf
   \usepackage[pdftex]{graphicx}
   \graphicspath{{./figures/}{./pdf/}{./jpeg/}}
   \DeclareGraphicsExtensions{.pdf,.jpeg,.png}
\else
   \usepackage[dvips]{graphicx}
   \graphicspath{{./eps/}}
   \DeclareGraphicsExtensions{.eps}
\fi
\usepackage{algorithm}
\usepackage{algpseudocode}

\usepackage[cmex10]{amsmath}
\usepackage{amssymb}

\usepackage{array}
\usepackage[caption=false,font=footnotesize]{subfig}
\usepackage{stfloats}

\usepackage{multirow}

\usepackage{footnote}

\usepackage{threeparttable}

\renewcommand{\vec}[1]{\mathbf{#1}}

\newcommand{\mtx}[1]{\mathbf{#1}}

\begin{document}
%
\title{Compression for Multiple Reconstructions}
%
%
%

\author{\IEEEauthorblockN{Yehuda Dar, Michael Elad, and Alfred M. Bruckstein}
	\IEEEauthorblockA{Computer Science Department, Technion -- Israel Institute of Technology}}

%
%

\markboth{}%
{~}
%



\maketitle

\begin{abstract}

In this work we propose a method for optimizing the lossy compression for a network of diverse reconstruction systems. 
We focus on adapting a standard image compression method to a set of candidate displays, presenting the decompressed signals to viewers.
Each display is modeled as a linear operator applied after decompression, and its probability to serve a network user.
We formulate a complicated operational rate-distortion optimization trading-off the network's expected mean-squared reconstruction error and the compression bit-cost. 
Using the alternating direction method of multipliers (ADMM) we develop an iterative procedure where the network structure is separated from the compression method, enabling the reliance on standard compression techniques.
We present experimental results showing our method to be the best approach for adjusting high bit-rate image compression (using the state-of-the-art HEVC standard) to a set of displays modeled as blur degradations.

\end{abstract}


~\\~
\begin{IEEEkeywords}
	Rate-distortion optimization, signal compression, image compression, image deblurring, alternating direction method of multipliers (ADMM).
\end{IEEEkeywords}

%
\IEEEpeerreviewmaketitle

\section{Introduction}
\label{sec:Introduction}
\footnotetext{Authors' E-mail addresses: \{ydar,~elad,~freddy\}@cs.technion.ac.il.}
Multimedia content is often distributed using broadcast and "on-demand" services reaching consumers with various display devices. Therefore, rendering the image/video can widely differ due to various technical aspects such as the specific display technology, different screen resolutions, etc. 
Such multimedia distribution systems fundamentally rely on lossy compression in order to meet storage and transmission-bandwidth limitations. 
However, while the displayed signals are the important outcomes of the flow, the compression is usually optimized only with respect to the decompressed signal, ignoring the subsequent processing and degradations occurring at the different displays.
In this work we study the problem of optimizing signal compression to a known set of display settings having different usage probabilities.


We recently \cite{dar2017optimized} proposed an optimization methodology to adjust standard image/video compression to a known type of display presenting the decompressed signal to the viewer. Our framework essentially pre-compensates the display degradation from the compression standpoint in a rate-distortion optimized manner.
Here we extend the problem settings of \cite{dar2017optimized} to optimize the compression with respect to a set of display devices, described by several linear rendering models and their probabilities to be in use by consumers. 
One can interpret the display models and their usage probabilities as a characterization of a multimedia distribution network.

We formulate a rate-distortion optimization to trade-off the compression bit-cost and the expected mean-squared error of the displayed signal. Similar to our previous works \cite{dar2017optimized,dar2018system}, we address the computationally hard optimization using the alternating direction method of multipliers (ADMM) \cite{boyd2011distributed} translating the task to sequentially applying standard compressions (that are network independent!) and $ \ell_2 $-constrained deconvolutions expressing the network structure. This procedure can be generically adapted to various network layouts and to any standard compression technique, providing network-optimized binary data that is compatible with desired standard decompression processes.

The problem settings we address here resemble the framework of (lossy) compression for computations applied on the decompressed data (see, e.g., \cite{misra2011distributed,sun2013distributed}). Yet, the post-decompression processing we consider here is mainly an unwanted degradation, in contrast to a desired computation. However, the ability of our method to adapt standard compression to post-decompression processing may be utilized also for compression tasks involving computations necessary to be carried out on the decompressed data.

The important problem of various display devices is treated also from the perspective of scalable image/video coding methods (see, e.g., the extension of the HEVC standard in \cite{boyce2016overview}), where the signal is coded in layers of increasing quality/resolution to be peeled by the network or the user device. In contrast, we take here another viewpoint on the problem, optimizing a single (non-layered) compression of a signal to a given collection of rendering models.

We demonstrate our general approach for image compression using the state-of-the-art HEVC standard coupled with various simplified display models in the form of linear blur operators following the decompression. 
While another recent method \cite{laparra2017perceptually} optimizes image rendering with respect to a perceptual quality metric, we present here (and in \cite{dar2017optimized}) a method to globally optimize the flow of compression, decompression and rendering. Since our optimization goal and the distortion type differ from those in \cite{laparra2017perceptually}, the two methods cannot be quantitatively compared. 
In our experiments we compared our approach to regular HEVC compression, and to preceding the compression with the Expected Patch Log Likelihood (EPLL) deblurring method \cite{zoran2011learning} adapted to the same fidelity term as we use in our method. The rate-distortion performance of the various methods clearly exhibit our method as the leading approach at high bit-rate compression.


\section{The Proposed Method}
\label{sec:Proposed Method}

\subsection{Problem Formulation}
\label{subsec:Problem Formulation}

\begin{figure*}[]
	\centering
	\includegraphics[width=0.85\textwidth]{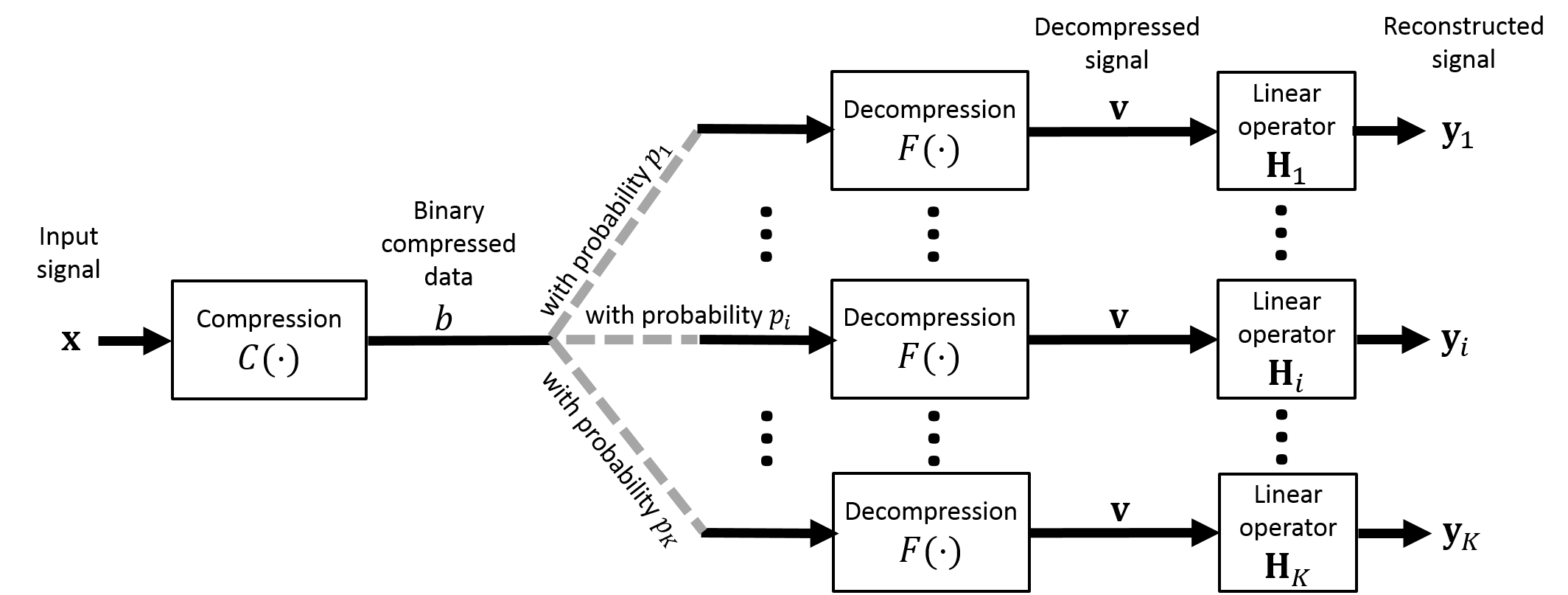}
	\caption{The general network structure considered in this paper.} 
	\label{Fig:general_problem_diagram}
\end{figure*}

We consider the network structure described in Fig. \ref{Fig:general_problem_diagram}, starting with an input signal in the form of an $ N $-length column-vector $ \vec{x} \in \mathbb{R}^N $ that is compressed and distributed over the network to users having various reconstruction systems. We describe the lossy compression procedure using the function $ C: \mathbb{R}^N \rightarrow \mathcal{B} $, mapping the $ N $-dimensional  input-signal domain to the discrete set $ \mathcal{B} $ of compressed representations in the form of variable-length binary descriptions. The compression of $ \vec{x} $ is denoted by $\textit{b} = C \left( \vec{x} \right)$, where $ \textit{b} \in \mathcal{B} $ is the compressed data to transmit over the network to an arbitrary number of users. 
The users have reconstruction systems that first decompress the data via $\vec{v} = F \left( \textit{b} \right)$, 
where $ F: \mathcal{B} \rightarrow \mathcal{S} $ maps the binary compressed representations in $ \mathcal{B} $ to the respective decompressed signals in the discrete set $ \mathcal{S} \subset \mathbb{R}^N $. The decompressed signal $ \vec{v} $ (which is an $ N $-length column vector) further goes through a linear operation, associated with a processing/degradation stage, that produces the reconstruction available to the user. In the case of visual signals, the post-decompression component may be a display rendering the viewed image. 
We assume a user may have one of $ K $ reconstruction systems (where $ K $ is a positive finite integer), differing in the linear operator applied after decompression. 
The post-decompression linear operator of the $ k^{th} $ system type ($ k=1,...,K $) is denoted by the $ N\times N $ matrix $ \mtx{H}_k $, producing a corresponding reconstructed (output) signal 
\begin{IEEEeqnarray}{rCl}
	\label{eq:network structure - reconstructed k^th output}
	\vec{y}_k  = \mtx{H}_k \vec{v} .
\end{IEEEeqnarray}

We assume the portions of using each of the $ K $ reconstruction systems are known and denoted by $ p_1, ..., p_K \ge 0 $, where $ \sum_{k=1}^{K} p_k = 1 $. Accordingly, a network user can be modeled to have a reconstruction system of a type corresponding to a discrete random variable over the values $ \left\lbrace 1, ..., K \right\rbrace $ with the respective probabilities $ p_1, ..., p_K $. Then, by (\ref{eq:network structure - reconstructed k^th output}) the reconstructed signal is a random vector $ \vec{y} $ having the value $ \vec{y}_k $ with probability $ p_k $ for $ k=1,...,K $. 
For a given (deterministic) input signal $ \vec{x} $ and its decompressed version $ \vec{v} $, and by the network structure, we quantify the expected mean-squared-error (MSE) of the reconstruction as 
\begin{IEEEeqnarray}{rCl}
	\label{eq:network structure - expected distortion}
	D\left( \vec{x}, \vec{v} \right) \triangleq \frac{1}{N} \sum\limits_{k=1}^{K} p_k \left\| { \vec{x} - \mtx{H}_k \vec{v} } \right\|_2^2 .
\end{IEEEeqnarray}

Our goal here is to optimize the rate-distortion performance of the network for a given input signal $ \vec{x} $. Accordingly, we formulate the task as the minimization of the compression bit-cost under constrained expected distortion (\ref{eq:network structure - expected distortion}), namely,  
\begin{IEEEeqnarray}{rCl}
	\label{eq:rate-distortion optimization - constrained}
	\begin{aligned}
		& \hat{ \vec{v}} = \underset{ \vec{v}\in\mathcal{S} }{\text{argmin}}
		& & { R \left( \vec{v} \right) } \\
		& \text{subject to}
		& & \frac{1}{N} \sum\limits_{k=1}^{K} p_k \left\| { \vec{x} - \mtx{H}_k \vec{v} } \right\|_2^2 \le d
	\end{aligned}
\end{IEEEeqnarray}
where $R \left( \vec{v} \right)$ evaluates the length of the binary compressed description $  \textit{b} \in \mathcal{B} $ matched to the decompressed signal $ \vec{v} $, and $ d \ge 0 $ is the allowed distortion.

Similar to contemporary compression tasks (see, e.g., \cite{ortega1998rate,sullivan2012overview}), we turn our optimization (\ref{eq:rate-distortion optimization - constrained}) into its unconstrained Lagrangian form 
\begin{IEEEeqnarray}{rCl}
	\label{eq:rate-distortion optimization - Lagrangian}
	\hat{ \vec{v}} = \underset{ \vec{v}\in\mathcal{S} }{\text{argmin}}
	~~ { R \left( \vec{v} \right) + \lambda \frac{1}{N} \sum\limits_{k=1}^{K} p_k \left\| { \vec{x} - \mtx{H}_k \vec{v} } \right\|_2^2 }
\end{IEEEeqnarray}
where $ \lambda \ge 0 $ is a Lagrange multiplier matching to a distortion constraint $ d_{\lambda} \ge 0 $ (such coding without a specified distortion constraint is prevalent, for instance, in video coding \cite{sullivan2012overview}).
Since we consider the compression of high-dimensional signals (i.e., $ N $ is large) the discrete set $ \mathcal{S} $ is prohibitively large, meaning that a direct solution of the Lagrangian form in (\ref{eq:rate-distortion optimization - Lagrangian}) is impractical for arbitrarily structured matrices $ \left\lbrace \mtx{H}_k \right\rbrace_{k=1}^{K} $. 
Note that when $ \mtx{H}_k = \mtx{I} $ for $ k =1,...,K$, the Lagrangian optimization in (\ref{eq:rate-distortion optimization - Lagrangian}) reduces to the standard compression form \cite{shoham1988efficient,ortega1998rate}, disregarding the network-oriented problem, and practically solvable using block-based architectures that decompose the problem to a sequence of block-level optimizations of sufficiently low dimensions.

\subsection{Practical Iterative Procedure}
\label{subsec:Practical Iterative Procedure}

We employ the alternating direction method of multipliers (ADMM) technique \cite{boyd2011distributed} to resolve the computationally challenging problem (\ref{eq:rate-distortion optimization - Lagrangian}) when the post-decompression operators $ \left\lbrace \mtx{H}_k \right\rbrace_{k=1}^{K} $ are arbitrarily structured.
We begin by splitting the optimization variable such that (\ref{eq:rate-distortion optimization - Lagrangian}) becomes 
\begin{IEEEeqnarray}{rCl}
	\label{eq:rate-distortion optimization - variable splitting}
	\begin{aligned}
		& \hat{ \vec{v}} = \underset{ \vec{v}\in\mathcal{S} , {\vec{z}}\in\mathbb{R}^N }{\text{argmin}}
		~~ { R \left( \vec{v} \right) + \lambda \frac{1}{N} \sum\limits_{k=1}^{K} p_k \left\| { \vec{x} - \mtx{H}_k \vec{z} } \right\|_2^2 } \\
		& \text{subject to} ~~~~ \vec{v} = \vec{z}
	\end{aligned}
\end{IEEEeqnarray}
where $ \vec{z} \in \mathbb{R}^N $ is an auxiliary variable that is not limited to the discrete set $ \mathcal{S} $.
Applying the scaled form of the augmented Lagrangian and the method of multipliers \cite[Ch. 2]{boyd2011distributed} on (\ref{eq:rate-distortion optimization - variable splitting}) yields an iterative process formulated as 
\begin{IEEEeqnarray}{rCl}
	\label{eq:rate-distortion optimization - augmented Lagrangian}
	&& \left( \hat{ \vec{v}}^{(t)}, \hat{\vec{z}}^{(t)} \right) = 
	\\ \nonumber
	&& \mathop {{\text{argmin}}}\limits_{ \vec{v}\in\mathcal{S} , {\vec{z}}\in\mathbb{R}^N }  R \left( \vec{v} \right) +  \frac{\lambda}{N} \sum\limits_{k=1}^{K} p_k \left\| { \vec{x} - \mtx{H}_k \vec{z} } \right\|_2^2 + \frac{\beta}{2}{\left\| { \vec{v} - \vec{z} + \vec{u}^{(t)} } \right\|_2^2} 
	\\ 
	&& \vec{u}^{(t+1)} = \vec{u}^{(t)} + \left( \hat{ \vec{v}}^{(t)} - \hat{\vec{z}}^{(t)} \right),
\end{IEEEeqnarray}
where $ t $ denotes the iteration index, $\vec{u}^{(t)} \in \mathbb{R}^N$ is the scaled dual variable, and $ \beta $ is an auxiliary parameter introduced by the augmented Lagrangian.
We get the ADMM form of the problem by applying one iteration of alternating minimization on (\ref{eq:rate-distortion optimization - augmented Lagrangian}), leading to the following sequence of easier optimizations 
\begin{IEEEeqnarray}{rCl}
	\label{eq:rate-distortion optimization - ADMM - compression}
	&& \hat{\vec{v}}^{(t)} = \mathop {{\text{argmin}}}\limits_{ \vec{v}\in\mathcal{S} }  R \left( \vec{v} \right) + \frac{\beta}{2}{\left\| {  \vec{v} - \tilde{ \vec{z}}^{(t)} } \right\|_2^2}
	\\
	\label{eq:rate-distortion optimization - ADMM - deconvolution}
	&& \hat{ \vec{z}}^{(t)} = \mathop {\text{argmin}}\limits_{{\vec{z}}\in\mathbb{R}^N } \lambda \frac{1}{N} \sum\limits_{k=1}^{K} p_k \left\| { \vec{x} - \mtx{H}_k \vec{z} } \right\|_2^2 + \frac{\beta}{2}{\left\| {  \vec{z} - \tilde{ \vec{v}}^{(t)} } \right\|_2^2} ~~~~~~
	\\
	\label{eq:rate-distortion optimization - ADMM - u update}
	&& \vec{u}^{(t+1)} = \vec{u}^{(t)} + \left( \hat{ \vec{v}}^{(t)} - \hat{\vec{z}}^{(t)} \right).
\end{IEEEeqnarray}
where $ \tilde{ \vec{z}}^{(t)} = \hat{\vec{z}}^{(t-1)} - \vec{u}^{(t)} $ and $ \tilde{ \vec{v}}^{(t)} = \hat{\vec{v}}^{(t)} + \vec{u}^{(t)} $. 
Nicely, the compression architecture $ \left\lbrace \mathcal{S}, R \right\rbrace $ and the network layout described by $ \left\lbrace \mtx{H}_k , p_k \right\rbrace_{k=1}^{K} $ were separated by the ADMM to distinct (and simpler) optimization tasks.

The optimization formulation in (\ref{eq:rate-distortion optimization - ADMM - compression}) coincides with the Lagrangian rate-distortion optimization utilized for standard compression tasks employing the usual (network independent) MSE distortion metric (here the effective Lagrange multiplier is $ \tilde{\lambda} = \frac{\beta N}{2 } $).
Hence, we propose to replace the solution of (\ref{eq:rate-distortion optimization - ADMM - compression}) with a standard compression (and decompression) method -- even one that does not exactly follow the Lagrangian optimization in (\ref{eq:rate-distortion optimization - ADMM - compression}).
We refer to the standard compression and decompression as 
\begin{IEEEeqnarray}{rCl}
	\label{eq:rate-distortion optimization - ADMM - compression - standard compression}
	{\textit{b}}^{(t)} = StandardCompress \left( \tilde{ \vec{z}}^{(t)}, \theta \right)
	\\
	\label{eq:rate-distortion optimization - ADMM - compression - standard decompression}
	\hat{\vec{v}}^{(t)} = StandardDecompress \left( {\textit{b}}^{(t)} \right)		
\end{IEEEeqnarray}
where $ \theta $ is a parameter generalizing the Lagrange multiplier part in regulating the rate-distortion tradeoff (see Algorithm \ref{Algorithm:Proposed Method}). The last generalizations establish the proposed procedure as a generic methodology for optimizing any compression method to particular network layouts.

The optimization in (\ref{eq:rate-distortion optimization - ADMM - deconvolution}) can be interpreted as an extended $ \ell_2 $-constrained deconvolution problem, here including a combination of several fidelity terms associated with the degradation operators $ \left\lbrace \mtx{H}_k \right\rbrace_{k=1}^{K} $. The analytic solution of (\ref{eq:rate-distortion optimization - ADMM - deconvolution}) is 
\begin{IEEEeqnarray}{rCl}
	\label{eq:rate-distortion optimization - ADMM - deconvolution - analytic solution}
	\nonumber \hat{ \vec{z}}^{(t)} = 
	\left(  \sum\limits_{k=1}^{K} {p_k \mtx{H}^{T}_{k}  \mtx{H}_{k} }  + \frac{\beta N}{2\lambda} \mtx{I}  \right)^{-1} \left(  \sum\limits_{k=1}^{K} {p_k \mtx{H}^{T}_{k} }  \vec{x} + \frac{\beta N}{2\lambda} \tilde{ \vec{v}}^{(t)}  \right)
\end{IEEEeqnarray}
exhibiting it as a linear combination of $ \vec{x} $ and $ \tilde{ \vec{v}}^{(t)}  $. 
We define the parameter  $ \tilde{\beta} \triangleq \frac{\beta N}{2\lambda} $ and use it in the generic method summarized in Algorithm \ref{Algorithm:Proposed Method}.

\begin{algorithm}
	\caption{Generic Network-Optimized Compression}
	\label{Algorithm:Proposed Method}
	\begin{algorithmic}[1]
		\State Inputs: $ \vec{x} $, $ \theta $, $ \tilde{\beta} $.
		\State  Initialize $t = 0$ , $ {\hat{\vec{z}}}^{(0)} = \vec{x} $ , $\vec{u}^{(1)} = \vec{0}$.
		\Repeat 
		
		\State $ t \gets t + 1$
		
		\State $ \tilde{ \vec{z}}^{(t)} = \hat{\vec{z}}^{(t-1)} - \vec{u}^{(t)} $
		
		\State $ {\textit{b}}^{(t)} = StandardCompress \left( \tilde{ \vec{z}}^{(t)}, \theta \right) $
		\State $ \hat{\vec{v}}^{(t)} = StandardDecompress \left( {\textit{b}}^{(t)} \right) $
		
		\State $ \tilde{ \vec{v}}^{(t)} = \hat{\vec{v}}^{(t)} + \vec{u}^{(t)} $
		\State $\hat{ \vec{z}}^{(t)} = \left(  \sum\limits_{k=1}^{K} {p_k \mtx{H}^{T}_{k}  \mtx{H}_{k} }  + \tilde{\beta} \mtx{I}  \right)^{-1} \left(  \sum\limits_{k=1}^{K} {p_k \mtx{H}^{T}_{k} }  \vec{x} + \tilde{\beta} \tilde{ \vec{v}}^{(t)}  \right)$
		
		\State $\vec{u}^{(t+1)} = \vec{u}^{(t)} + \left( \hat{ \vec{v}}^{(t)} - \hat{\vec{z}}^{(t)} \right)$
		
		\Until{stopping criterion is satisfied}
		\State Output: $ {\textit{b}}^{(t)} $, which is the binary compressed data obtained in the last iteration.
	\end{algorithmic}
\end{algorithm}

\section{Experimental Results}
\label{sec:Experimental Results}

\begin{figure*}[]
	\centering
	{\subfloat[Regular Decompressed: 1.65 bpp]{\label{fig:flowerbug_regular_decompressed_1_65bpp}\includegraphics[width=0.24\textwidth]{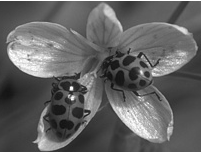}}}
	{\subfloat[Regular: Display 1 (38.42dB)]{\label{fig:flowerbug_regular_decompressed_1_65bpp__blurred1_psnr_38_42dB}\includegraphics[width=0.24\textwidth]{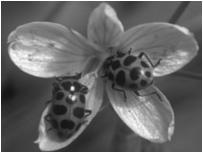}}}
	{\subfloat[Regular: Display 2 (35.11 dB)]{\label{fig:flowerbug_regular_decompressed_1_65bpp__blurred2_psnr_35_11dB}\includegraphics[width=0.24\textwidth]{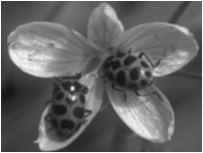}}}
	{\subfloat[Regular: Display 3 (33.28 dB)]{\label{fig:flowerbug_regular_decompressed_1_65bpp__blurred3_psnr_33_28dB}\includegraphics[width=0.24\textwidth]{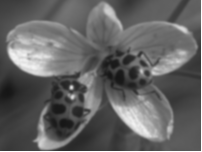}}}
	\\
	{\subfloat[Proposed Decompressed: 1.61 bpp]{\label{fig:flowerbug_our_decompressed_1_61bpp}\includegraphics[width=0.24\textwidth]{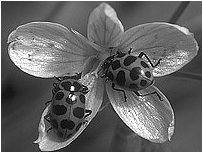}}}
	{\subfloat[Proposed: Display 1 (46.60 dB)]{\label{fig:flowerbug_our_decompressed_1_61bpp__blurred1_psnr_46_60dB}\includegraphics[width=0.24\textwidth]{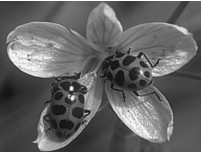}}}
	{\subfloat[Proposed: Display 2 (39.52 dB)]{\label{fig:flowerbug_our_decompressed_1_61bpp__blurred2_psnr_39_52dB}\includegraphics[width=0.24\textwidth]{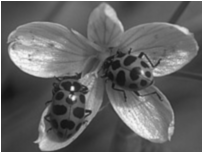}}}
	{\subfloat[Proposed: Display 3 (35.38 dB)]{\label{fig:flowerbug_our_decompressed_1_61bpp__blurred3_psnr_35_38dB}\includegraphics[width=0.24\textwidth]{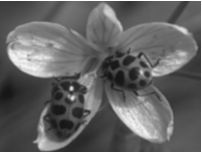}}}
	\caption{The regular and the proposed method applied for the 'Flower and Bugs' image. {The denoted PSNR values in this figure are for the individual reconstructions, i.e., using the regular MSE and not the expected one from (\ref{eq:network structure - expected distortion}) that is used in the rest of the paper.} } 
	\label{Fig:Experiments - blur - flowerbug}
\end{figure*}

Let us demonstrate our method for optimizing the HEVC still-image compression standard (implemented in the software in \cite{hevc_software_bpg}) to three possible blur operators degrading the decompressed image. 
The post-decompression linear operators $ \mtx{H}_1 $, $ \mtx{H}_2 $, and $ \mtx{H}_3 $ correspond to shift-invariant Gaussian blur kernels (of $ 15\times 15 $ pixels size) having standard deviations 0.6, 0.8, and 1, respectively, and usage probabilities of $ p_1 \nolinebreak= \nolinebreak 0.6 $, $ p_2 = 0.3 $, and $ p_3 = 0.1 $.

\begin{figure}
	\centering
	{\subfloat[Segment of the Input Image]{\label{fig:flowerbug_input}\includegraphics[width=0.24\textwidth]{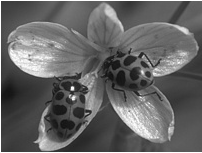}}}~~
	{\subfloat[PSNR-bitrate curves]{\label{fig:berkeley_flower_and_bugs_RD_curves}\includegraphics[width=0.23\textwidth]{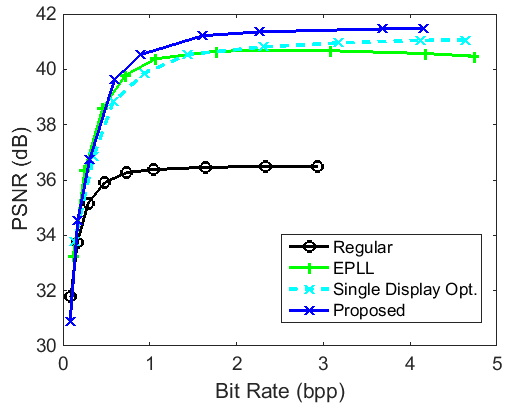}}}
	\caption{Methods evaluation for the image 'Flower and Bugs'.} 
	\label{Fig:Experimental Results - PSNR-rate curves}
\end{figure}

To evaluate our method we constructed three competing techniques also using HEVC image compression, and compared them to our method\footnote{The Peak Signal-to-Noise Ratio (PSNR) here relies on the expected reconstruction MSE given in (\ref{eq:network structure - expected distortion}), i.e., $PSNR = 10 \log_{10} \left( {P^2}/{D\left( \vec{x}, \vec{v} \right)} \right)$ where $\vec{x}$ and $\vec{v}$ are the input and the decompressed signals, respectively, and $ P $ is the maximal signal-value generally possible.}. The PSNR-bitrate curves of the examined methods (see, e.g., Fig. \ref{fig:berkeley_flower_and_bugs_RD_curves}) were created for each of the 12 examined images (see Table \ref{table:Experiments - blur - Average PSNR and Bit-Rate comparison}) by applying their HEVC component for 9 quality parameter (QP) values equally-spaced between 1 to 41. 
The first competing approach is to regularly compress without any pre/post processing (while the decompression is still followed by the inevitable deterioration). As expected, this naive method performs poorly.
The second competing procedure precedes the compression with deconvolution using the Expected Patch Log Likelihood (EPLL) method relying on a Gaussian Mixture Model (GMM) prior learned for natural images (see \cite{zoran2011learning}). The EPLL implementation used here is with respect to a fidelity term corresponding to (\ref{eq:network structure - expected distortion}) and suitable parameter settings.
The third competing method is our pre-compensating compression from \cite{dar2017optimized}, optimized only for a single display (corresponding to the highest probability).

The implementation of the proposed method (Algorithm \ref{Algorithm:Proposed Method}) uses a $ \tilde{\beta} $ value based on the HEVC quality parameter (the $ \tilde{\beta} $ value here is 10 times the value formulated in \cite{dar2017optimized}). The stopping criterion was defined to a maximal number of 40 iterations or to end earlier when $ \hat{\vec{v}}^{(t)} $ and $ \hat{ \vec{z}}^{(t)} $ converge or diverge (as described in \cite{dar2017optimized}).

The evaluation of the PSNR-bitrate curves summarized in Table \ref{table:Experiments - blur - Average PSNR and Bit-Rate comparison} and exemplified for one image in Fig. \ref{fig:berkeley_flower_and_bugs_RD_curves}, showing that our method outperforms the other techniques at high bit-rates, where we achieve significant PSNR gains compared to the regular, the EPLL-based, and the single display optimization procedures. The average PSNR gains in Table \ref{table:Experiments - blur - Average PSNR and Bit-Rate comparison} were computed based on the BD-PSNR metric \cite{bjontegaard2001calculation,BDPSNR_Matlab} for the high bit-rate range (here defined by QP values 1,6,11,16).

In Figure \ref{Fig:Experiments - blur - flowerbug} we present visual results for the compression of the 'Flower and Bugs' image (see Fig. \ref{fig:flowerbug_input}, where only a portion of the input image is presented due to lack of space).  Figures \ref{fig:flowerbug_regular_decompressed_1_65bpp} and \ref{fig:flowerbug_our_decompressed_1_61bpp} exhibit the decompressed images (before degradation) using the regular approach and the proposed method, respectively. Figures \ref{fig:flowerbug_regular_decompressed_1_65bpp__blurred1_psnr_38_42dB}-\ref{fig:flowerbug_regular_decompressed_1_65bpp__blurred3_psnr_33_28dB} and \ref{fig:flowerbug_our_decompressed_1_61bpp__blurred1_psnr_46_60dB}-\ref{fig:flowerbug_our_decompressed_1_61bpp__blurred3_psnr_35_38dB} show the three simulated displayed versions of the decompressed images. Evidently, we get significantly higher PSNR values (that correspond to the regular MSE measure) at a similar (slightly lower) bit-rate. Our method produces an overly-sharpened decompressed image (Fig. \ref{fig:flowerbug_our_decompressed_1_61bpp}) that is later balanced with the rendering blur, leading to better displayed images (Figs. \ref{fig:flowerbug_our_decompressed_1_61bpp__blurred1_psnr_46_60dB}-\ref{fig:flowerbug_our_decompressed_1_61bpp__blurred3_psnr_35_38dB}).

\begin{table} []
	\caption{Method Evaluation for 12 Images}
	\renewcommand{\arraystretch}{1.1}
	\label{table:Experiments - blur - Average PSNR and Bit-Rate comparison}
	\centering
	\begin{tabular}{|m{1cm}|c||c|c|c|}
		\hline
		\multirow{2}{*}{\bfseries \shortstack{Dataset}}  & \multirow{2}{*}{\bfseries \shortstack{Image}}  & \multicolumn{3}{|c|}{\bfseries \shortstack{\\Average PSNR Gains at High Bit-Rates}} \\
		\cline{3-5}
		& & \shortstack{\\Proposed\\over\\Regular} & \shortstack{\\Proposed\\over\\EPLL} & \shortstack{\\Proposed\\over\\Single Display} \\
		\hline
		\hline				                                       
		\cite{asuni2014testimages} & Almonds  & 5.25 & 0.47 & 0.49 \\
		\cline{2-5}	
		TEST & Cards  & 4.83 & 0.56 & 0.35 \\
		\cline{2-5}	
		IMAGES  & Duck toys  & 4.83 & 0.56 & 0.49 \\
		\cline{2-5}	
		300x300 & Garden table  & 4.70 & 0.48 & 0.35 \\
		\hline
		\hline
		\cite{schaefer2003ucid}& House \& lawn  & 3.53 & 1.36 & 0.25 \\
		\cline{2-5}	
		UCID & Tree  & 3.83 & 1.18 & 0.20 \\
		\cline{2-5}	
		384x512 & Garden  & 3.26 & 1.44 & 0.17 \\
		\cline{2-5}	
		& Teddy bear  & 4.96 & 0.73 & 0.48 \\
		\hline
		\hline				                                       
		\cite{martin2001database}& Bears & 4.73 & 0.64 & 0.35 \\
		\cline{2-5}	
		Berkeley  & Boats & 4.05 & 0.90 & 0.28 \\ 
		\cline{2-5}	
		481x321 & Butterfly & 4.85 & 0.67 & 0.39 \\
		\cline{2-5}	
		& Flower \& Bugs & 4.83 & 0.74 & 0.49 \\
		\hline		                                               
	\end{tabular}
\end{table}

\section{Conclusion}
\label{sec:Conclusion}


We presented a method for optimizing compression to a set of reconstruction systems, each has a different linear processing/degradation after decompression. 
Using ADMM we established a generic compression procedure relying on a standard compression technique. 
Experiments for adjusting image compression (using the HEVC standard) to a set of blur operators modeling display degradations showed our method as the leading approach for high bit-rate compression.

%
%

\ifCLASSOPTIONcaptionsoff
  \newpage
\fi



\bibliographystyle{IEEEtran}
\bibliography{IEEEabrv,compression_for_multiple_displays_conference__refs}

\begin{thebibliography}{10}
\providecommand{\url}[1]{#1}
\csname url@samestyle\endcsname
\providecommand{\newblock}{\relax}
\providecommand{\bibinfo}[2]{#2}
\providecommand{\BIBentrySTDinterwordspacing}{\spaceskip=0pt\relax}
\providecommand{\BIBentryALTinterwordstretchfactor}{4}
\providecommand{\BIBentryALTinterwordspacing}{\spaceskip=\fontdimen2\font plus
\BIBentryALTinterwordstretchfactor\fontdimen3\font minus
  \fontdimen4\font\relax}
\providecommand{\BIBforeignlanguage}[2]{{%
\expandafter\ifx\csname l@#1\endcsname\relax
\typeout{** WARNING: IEEEtran.bst: No hyphenation pattern has been}%
\typeout{** loaded for the language `#1'. Using the pattern for}%
\typeout{** the default language instead.}%
\else
\language=\csname l@#1\endcsname
\fi
#2}}
\providecommand{\BIBdecl}{\relax}
\BIBdecl

\bibitem{dar2017optimized}
Y.~Dar, M.~Elad, and A.~M. Bruckstein, ``Optimized pre-compensating
  compression,'' \emph{Submitted to IEEE Trans. Image Process., arXiv preprint
  arXiv:1711.07901}, 2017.

\bibitem{dar2018system}
------, ``System-aware compression,'' \emph{arXiv preprint arXiv:1801.04853},
  2018.

\bibitem{boyd2011distributed}
S.~Boyd, N.~Parikh, E.~Chu, B.~Peleato, and J.~Eckstein, ``Distributed
  optimization and statistical learning via the alternating direction method of
  multipliers,'' \emph{Foundations and Trends in Machine Learning}, vol.~3,
  no.~1, pp. 1--122, 2011.

\bibitem{misra2011distributed}
V.~Misra, V.~K. Goyal, and L.~R. Varshney, ``Distributed scalar quantization
  for computing: High-resolution analysis and extensions,'' \emph{IEEE Trans.
  Inf. Theory}, vol.~57, no.~8, pp. 5298--5325, 2011.

\bibitem{sun2013distributed}
J.~Z. Sun, V.~Misra, and V.~K. Goyal, ``Distributed functional scalar
  quantization simplified,'' \emph{IEEE Trans. Signal Process.}, vol.~61,
  no.~14, pp. 3495--3508, 2013.

\bibitem{boyce2016overview}
J.~M. Boyce, Y.~Ye, J.~Chen, and A.~K. Ramasubramonian, ``Overview of {SHVC}:
  Scalable extensions of the high efficiency video coding standard,''
  \emph{IEEE Trans. Circuits and Systems for Video Technology}, vol.~26, no.~1,
  pp. 20--34, 2016.

\bibitem{laparra2017perceptually}
V.~Laparra, A.~Berardino, J.~Ball{\'e}, and E.~P. Simoncelli, ``Perceptually
  optimized image rendering,'' \emph{Journal of the Optical Society of America
  A}, vol.~34, p. 1511, 2017.

\bibitem{zoran2011learning}
D.~Zoran and Y.~Weiss, ``From learning models of natural image patches to whole
  image restoration,'' in \emph{IEEE ICCV}, 2011, pp. 479--486.

\bibitem{ortega1998rate}
A.~Ortega and K.~Ramchandran, ``Rate-distortion methods for image and video
  compression,'' \emph{IEEE Signal Process. Mag.}, vol.~15, no.~6, pp. 23--50,
  1998.

\bibitem{sullivan2012overview}
G.~J. Sullivan, J.~Ohm, W.-J. Han, and T.~Wiegand, ``Overview of the high
  efficiency video coding (hevc) standard,'' \emph{IEEE Trans. Circuits and
  Systems for Video Technology}, vol.~22, no.~12, pp. 1649--1668, 2012.

\bibitem{shoham1988efficient}
Y.~Shoham and A.~Gersho, ``Efficient bit allocation for an arbitrary set of
  quantizers,'' \emph{IEEE Trans. Acoust., Speech, Signal Process.}, vol.~36,
  no.~9, pp. 1445--1453, 1988.

\bibitem{hevc_software_bpg}
\BIBentryALTinterwordspacing
F.~Bellard, ``{BPG} 0.9.6.'' [Online]. Available: \url{http://bellard.org/bpg/}
\BIBentrySTDinterwordspacing

\bibitem{bjontegaard2001calculation}
G.~Bjontegaard, ``Calculation of average {PSNR} differences between
  {RD}-curves,'' in \emph{ITU-T Q. 6/SG16 VCEG, 15th Meeting, Austin, Texas,
  USA, April, 2001}.

\bibitem{BDPSNR_Matlab}
\BIBentryALTinterwordspacing
G.~Valenzise, ``Bjontegaard metric ({Matlab} function).'' [Online]. Available:
  \url{http://www.mathworks.com/matlabcentral/fileexchange/27798-bjontegaard-metric}
\BIBentrySTDinterwordspacing

\bibitem{asuni2014testimages}
N.~Asuni and A.~Giachetti, ``{TESTIMAGES}: a large-scale archive for testing
  visual devices and basic image processing algorithms.'' in \emph{Eurographics
  Italian Chapter Conference}, 2014, pp. 63--70.

\bibitem{schaefer2003ucid}
G.~Schaefer and M.~Stich, ``{UCID}: an uncompressed color image database,'' in
  \emph{Storage and Retrieval Methods and Applications for Multimedia}, vol.
  5307.\hskip 1em plus 0.5em minus 0.4em\relax International Society for Optics
  and Photonics, 2003, pp. 472--481.

\bibitem{martin2001database}
D.~Martin, C.~Fowlkes, D.~Tal, and J.~Malik, ``A database of human segmented
  natural images and its application to evaluating segmentation algorithms and
  measuring ecological statistics,'' in \emph{IEEE ICCV}, 2001, pp. 416--423.

\end{thebibliography}
%

%
%

%






\end{document}